# Close encounters in a pediatric ward: measuring face-to-face proximity and mixing patterns with wearable sensors

L. Isella[1], M. Romano[2], A. Barrat[1,3], C. Cattuto[1], V. Colizza[1], W. Van den Broeck[1], F. Gesualdo[2] E. Pandolfi[2], L. Ravà[2], C. Rizzo[4], A.E. Tozzi[2,*]

[1]Complex Networks and Systems Group, Institute for Scientific Interchange (ISI) Foundation, Torino, Italy
[2]Epidemiology Unit, Bambino Gesù Hospital, Rome, Italy
[3]Centre de Physique Théorique de Marseille, CNRS UMR 6207, Marseille, France
[4]National Centre for Epidemiology, Surveillance and Health Promotion, Istituto Superiore di Sanità Rome, Italy

[*]Corresponding author: Alberto E Tozzi, alberto.tozzi@gmail.com

**Abstract**

Background

Nosocomial infections place a substantial burden on health care systems and represent one of the major issues in current public health, requiring notable efforts for its prevention. Understanding the dynamics of infection transmission in a hospital setting is essential for tailoring interventions and predicting the spread among individuals. Mathematical models need to be informed with accurate data on contacts among individuals.

Methods and Findings

We used wearable active Radio-Frequency Identification Devices (RFID) to detect face-to-face contacts among individuals with a spatial resolution of about 1.5 meters, and a time resolution of 20 seconds. The study was conducted in a general pediatrics hospital ward, during a one-week period, and included 119 participants, with 51 health care workers, 37 patients, and 31 caregivers. Nearly 16,000 contacts were recorded during the study period, with a median of approximately 20 contacts per participants per day. Overall, 25% of the contacts involved a ward assistant, 23% a nurse, 22% a patient, 22% a caregiver, and 8% a physician. The majority of contacts were of brief duration, but long and frequent contacts especially between patients and caregivers were also found. In the setting under study, caregivers do not represent a significant potential for infection spread to a large number of individuals, as their interactions mainly involve the corresponding patient. Nurses would deserve priority in prevention strategies due to their central role in the potential propagation paths of infections.

Conclusions

Our study shows the feasibility of accurate and reproducible measures of the pattern of contacts in a hospital setting. The obtained results are particularly useful for the study of the spread of respiratory infections, for monitoring critical patterns, and for setting up tailored prevention strategies. Proximity-sensing technology should be considered as a valuable tool for measuring such patterns and evaluating nosocomial prevention strategies in specific settings.

**Introduction**

The knowledge of contact patterns among individuals is of paramount importance in the control of infections transmitted from person to person. Patterns of transmission of infections in a given population can be studied using data on contact patterns, knowing the transmission route of the infection, the characteristics of the infectious agent, and the immunity profiles of individuals. The knowledge of patterns of transmission is essential to identify specific mechanisms that favor transmission and thus to set up tailored intervention strategies such as social or physical barriers, targeted immunization, pharmaceutical interventions, and other measures aimed at preventing transmission and controlling the spread of the disease (1-8).

The parameters relevant to contact patterns are mostly obtained from surveys conducted in samples of individuals or can be indirectly estimated from seroprevalence data or from time-use data, and are then extrapolated to the general population (9-15). Several empirical studies have been recently conducted to determine the pattern of contacts between and within groups in different social settings to model the spread of infectious diseases, most often relying on self-reported methods (10, 16). The study of contacts through interviews and recall of previous encounters has however some limitations, as the data collection is not based on objective measurements and is generally performed on a random day, thus lacking the longitudinal dimension (17). Most importantly, this approach has a limited applicability to specific settings, such as, e.g., hospitals, that require high-resolution information both at the spatial and temporal level to accurately characterize the interactions among individuals, in an objective way and by means of non-obtrusive methodologies.

The lack of such information has so far limited the guidelines on preventive strategies for the epidemiology of hospital acquired infections mostly to general and qualitative recommendations, that cannot take into account the characteristics of patients, procedures, hospital wards, structures and logistics, and the related heterogeneities across diverse hospital settings (18). Similarly, while mathematical models have begun to provide insights into many important questions in the epidemiology of nosocomial infections (19), these results mostly focus on qualitative predictions (20), due to the very basic and homogeneous approaches adopted in the description of the population contact structure. On the other hand, fine-grained behavioral information may be used to inform and validate agent-based models of nosocomial infection (35).

To overcome such limitations and address the lack of empirical data relevant to the study of nosocomial infections, we rely on the technology of networked wearable sensors, that provides a novel approach to obtain accurate and tailored estimates of the contact patterns at the individual level in a given community (21-25,36). Specifically, we use active Radio-Frequency Identification Devices (RFID) as wearable sensors to measure face-to-face proximity with a high spatial and temporal resolution (23-25). The deployment of these devices may support the collection of accurate and timely information useful to inform and parameterize models for the study and prediction of the transmission of nosocomial infections, as well as to tailor containment measures for the control of potential outbreaks. To this aim, we present here the results of a study on the measurement and analysis of contact patterns within a pediatric hospital involving patients, caregivers, and health care workers. Our work illustrates the feasibility of contact measures through RFID devices in hospital settings, aimed at providing data-driven knowledge to inform models and prevention strategies.

**Methods**

Study setting

The study was performed in a general pediatric ward of the Bambino Gesù Hospital in Rome, Italy, a large third level pediatric hospital. The ward under study has 44 beds arranged in 22 rooms with 2 beds each, and mostly admits children with acute diseases who do not require intensive care or surgery. The occupancy rate of the ward is rarely below 95%. The pediatric ward is located in the Department of Pediatrics and is physically separated from other wards and facilities of the Hospital. In a typical workday 10 physicians, 20 nurses, and 6 ward assistants are on duty. Patients admitted into the ward are accompanied by one parent or tutor who spends the night in the same room on a chair. Visitors are allowed to visit the ward during the scheduled visit time of 1 hour in the afternoon. The study was conducted from the 9th to the 16th of November 2009 (weeks 46 and 47), during the incidence peak of the influenza A/H1N1v pandemic in Italy (25).

Data collection infrastructure

Data were obtained and collected through the deployment of an infrastructure consisting of a distributed sensing component, comprising wearable active Radio-Frequency Identification Devices (RFID), and of a data collection and processing component made of RFID readers installed in the ward, a local area network (LAN), and a central computer system for data collection and storage. The deployed RFID devices exchange ultra-low power radio packets in a peer-to-peer fashion. The devices perform a scan of their neighborhood by alternating transmit and receive cycles. During the transmit phase, low-power packets are sent out on a specific radio channel; during the receive phase, the devices listen on the same channel for packets sent by nearby devices. By including the transmitting signal strength in the payload, the receiving device can estimate the degree of proximity of the transmitting device, and this operation can be carried out in a decentralized fashion throughout the sensing network. A more detailed description of the data gathering infrastructure is reported in Refs. (23-25).

The lowest power level used in this deployment was selected to allow packet exchange only between devices within 1–1.5 meters of one another. This setting ensures that when individuals wear the devices on their chest, exchange of radio packets between RFID devices is only possible when they are facing each other, as the human body acts as a RF shield at the carrier frequency used for communication. When a relation of face-to-face proximity is detected, it is relayed from the RFID devices to the RFID readers installed in the hospital ward. The receiving infrastructure only covered the area within the hospital ward under study. The RFID devices are embedded in small hermetically sealed badges to comply with the security regulations of the hospital, and to minimize potential damage arising from their use. Each device has a unique identification number that was used to link the information on the contacts established by the person carrying the device with his/her profile.

Study design

The study was approved by the Ethical Committee of the Bambino Gesù Hospital. Before the study started, health care workers were invited to participate in a meeting where the project was fully explained and its procedure was illustrated in detail. Health care personnel, patients, and their caregivers (defined as tutors, accompanying persons and visitors) were systematically invited to participate in the study. At the time of enrollment, all participants signed an informed consent form, were given an RFID badge, and were asked to wear it at all times within the ward. The RFID tags of a patient and accompanying persons were returned to the study personnel at patient discharge or when moved to another ward. Visitors returned their tag at the end of the visit, whereas health care workers wore RFID tags for the entire duration of the study during working hours. Correct wearing of the RFID tags was monitored daily, and in case of loss or other anomalies the tag was replaced with a new one. The correct operation of devices was also monitored by reviewing the quality of the received signals on a daily basis, and replacing malfunctioning devices. While patients were fully identified to link personal information to pattern of contacts detected by RFID devices, no personal identifiers were used for caregivers or health care workers participating in the study, which were only associated to their professional category. However, the association between a given health care worker and the corresponding RFID tag was fixed throughout the study, i.e., it was possible to track the behavior of the (anonymous) given individual across hospital shifts and across working days. Overall, individuals were distinguished into the following classes or roles for the purpose of the present study: ward assistants (A), who are health care workers in charge of cleaning the ward and distributing the meals, physicians (D), nurses (N), patients (P), and caregivers (C), who comprise tutors and non professional visitors.

Data processing and analysis

We tuned the rate at which low-power packets are emitted and the fraction of time the devices spend listening on the contact channel so that the face-to-face proximity of two individuals wearing the RFID tags can be assessed with a probability in excess of 99% over an interval of 20 seconds. This sets the time scale over which we perform the temporal aggregation of the collected data, allowing for an adequate description of person-to-person interactions that includes brief encounters. Faster timescales are prone to increasing noise and do not result in a higher accuracy of the detection process. We defined that a 'contact' occurs between two individuals during an interval of 20s if and only if the RFID devices worn by the individuals exchanged at least one packet at the lowest power level during that interval. After a contact is established, it is considered ongoing as long as the devices continue to exchange at least one such packet for every subsequent 20s interval. Conversely, a contact is considered broken if a 20s interval elapses with no exchange of low-power packets.

The patterns of close encounters between individuals and the mixing patterns among classes of individuals were analyzed through several indicators. The high spatial and temporal resolution of the infrastructure allows us to monitor the number of contacts that each individual establishes with any other individual, to record the time spent on each such encounter, the cumulative time spent in contact between two individuals, and the frequency of encounters between any two individuals. More in detail, for each pair of individuals $i$ and $j$ we define several possible weights $w_{ij}$, each corresponding to a different quantity measured on the collected data: the *occurrence of the contact* $w_{ij}^p$, with $w_{ij}^p = 1$ if at least one contact between $i$ and $j$ has been established, and 0 otherwise; the *frequency of the contact* $w_{ij}^n$, indicating how many times the contact between $i$ and $j$ is observed during the study; the time spent on each such encounter; the *cumulative duration of the contact* $w_{ij}^t$, indicating the sum of the durations of all contacts established between $i$ and $j$ observed during the study. In addition to the above quantities, that are weights defined for pairs of individuals $i$ and $j$, it is possible to define corresponding quantities $s_i$ for each individual $i$, aggregating on all individuals $j$ who had a contact with $i$, i.e. $s_i = \sum w_{ij}$. In relation to the previously defined weights, one obtains the following quantities: the *number of distinct contacts* $s_i^p$, indicating the number of distinct individuals with whom $i$ has established at least one contact (i.e., a contact between $i$ and $j$ that occurs $w_{ij}^n > 1$ times is counted only once); the *number of contacts* $s_i^n$, indicating the overall number of contacts established by individual $i$, counting repeated contacts with the same individual $j$ as distinct events; the *cumulative time in contact* $s_i^t$, corresponding to the total sum of the duration of all contacts involving individual $i$.

In the following, unless otherwise stated, all measures are normalized on a 24-hour (daily) interval, e.g., all cumulative times in contact are divided by the duration of the study expressed in days (8 days). Figure 1 provides an example on how the above quantities are computed for a schematic sequence of contact data, where the individuals $i$ and $j$ belong to two different classes, nurses (N) and patients (P).

**Results**

Sample characteristics

A total of 195 individuals were invited to participate in the study, and 7 of them (3.6%, 5 health care workers, 1 patient, 1 caregiver) declined participation. A total of 188 individuals were provided with an RFID tag. Before performing the analysis, we reviewed the collected data with specific regard to the quality of signals, and monitored the continuous detection of signals from the devices. In particular, for patients and caregivers classes, we compared the duration of their physical presence at the ward, as obtained from hospital records, with the presence recorded from the RFID infrastructure. We identified and excluded 69 RFIDs belonging to patients or caregivers whose signals were not continuously detected by the receiving infrastructure and were missing for at least 25% of the time they were assigned to the individuals. Given that patients and caregivers spend most of the time at a fixed location in the ward, the above exclusion criterion mainly has the purpose of ensuring good radio coverage of the corresponding rooms. No exclusion was considered in the case of health care workers, as their RFID tags were anonymized and therefore not traceable to their duties; no comparison with paper records was possible, and all signals from those tags were thus retained in the analysis. Results were statistically assessed in terms of their robustness with respect to such filtering procedure and to the chosen 25% threshold, and no strong dependence on the selected subset of individuals was found, as shown in Fig. S1 of the Supplementary Information, which displays unfiltered data.

The analysis therefore included a total of 119 individuals: 10 ward assistants (8.4% of the total number of individuals), 20 physicians (16.8%), 21 nurses (17.6%), 37 patients (31.1%) and 31 caregivers (26.1%). Table 1 reports the characteristics of patients and their discharge diagnoses with regard to airborne infections. Age, length of stay, and the proportion of children with any respiratory infections were similar when comparing the two subsets of patients retained and excluded from the study because of the filtering procedure. We used the median test for comparing age and length of stay, and the chi-square test for the distribution of diagnoses, and the statistical comparison did not yield results signaling any significant difference between the two subsets. Despite the majority of patients was discharged with a diagnosis of a potentially transmissible disease, during the study period no evidence of transmission of nosocomial infection in patients or health care workers was reported according to the routine nosocomial infection surveillance system.

Number, frequency and duration of contacts by category

A total of nearly 16,000 contacts were recorded during the study period, with a median of approximately 20 contacts per participants per day. Table 2 reports the total number of contacts established by individuals for each class, and the median values per participant of the given category. A very large number of contacts involve at least a ward assistant or a nurse. In addition, significant fractions of the overall number of contacts are found to involve at least a patient or a caregiver, representing approximately 22% of the total in both cases. The minimum number of contacts is observed for physicians (8%).

The daily median number of contacts per individual, shown in Figure 2A, is highest among ward assistants, and it is much higher than the values observed for other classes (the second largest value is 45.5 contacts established by an average nurse, compared to 99.5 contacts in the ward assistant case). This pattern is explained by the organization of the staff duties, as ward assistants are requested mainly to clean up, to transport patients, and to distribute food, and they mostly work in pairs. Figure 2B reports the daily median number of distinct contacts $s^p$ that participants of a given class establish with any other participant, showing that health-care workers are involved in interactions with a larger number of distinct individuals, typically 3 – 5 distinct contacts, whereas patients and caregivers mostly interact with a single person, with high frequency. In addition to the number and the frequency of contacts, the RFID infrastructure enables the accurate measurement of the duration of contacts between individuals, providing a more detailed characterization of proximity events, which is generally hard to achieve by using traditional survey methodologies. Figure 2C shows the cumulative time in contact, $s^t$, spent by individuals belonging to each of the roles. This quantity is highly heterogeneous, ranging, daily, from a few minutes for physicians up to approximately one hour for ward assistants.

The computed quantities show large heterogeneities among individuals of the same class and across the various days of the study. Figure 3A reports the probability density function $P(s^n)$ (number of events in each bin divided by the bin width) of the number of contacts $s^n$ obtained for each class of participants. The distribution $P(s^n)$ for a given class is defined as the probability that a randomly selected participant of that class has established a total of $s^n$ contacts with any other individual during a given day. Large fluctuations are visible in the number of contacts per individual, varying over 2 or 3 orders of magnitude, and the largest fluctuations are observed in the case of patients and caregivers. Figure 3B reports the probability density functions of the cumulative time in contact, for individuals of each class. The longest durations, up to nearly 4 hours, are observed for contacts that involve at least one patient or one caregiver. The observed broad probability distributions are typical of human-driven systems and have been already observed elsewhere (24,25). Figure 4 shows boxplots for the distributions of cumulative contact durations $w^t$ between pairs of individuals belonging to given categories. Overall, about 95% of the contacts have a cumulated duration $w^t$ of less than 4 minutes. Besides characterizing the overall behavior of a given class of participants interacting with any other individual, the collected data allow to further inspect the interaction behavior by focusing on face-to-face proximity between pairs of categories. Figure 5 reports three contact matrices defined on the classes of participants, taking into account the different numbers of individuals in each class (3), and measured using the three quantities defined above: the number of contacts $s^n$ (panel A), the number of distinct contacts $s^p$ (panel B), and the cumulative time in contact $s^t$ (panel C). Panel A shows that the majority of contacts occur within the ward assistant class, followed by nurse-nurse interactions. A number of contacts larger than 10 is also observed for patient-caregiver interactions, considering both the number of contacts that a patient had with any caregiver, and the number of contacts that a caregiver had with any patient. These contacts, however, are characterized by a very high frequency, as signaled by the very small number of distinct contacts reported in panel B for the same two classes, consistent with a strong one-to-one patient-caregiver interaction. A smaller number of distinct contacts is also observed among ward assistants, compared to the median value of 63 contacts, whereas the nurse-nurse interaction remains strong also in terms of number of distinct contacts. The number of contacts (both distinct and non-distinct) among patients is noticeably very low and close to zero. The contact matrix computed in terms of the cumulative time in contact provides yet another characterization of the interaction behavior among classes. Long interactions are observed between patients and caregivers, among nurses, and among ward assistants. The time spent in close proximity by a pair of patients or by a pair of visitors is extremely small, and interactions between a health care worker and a non-health care worker are very limited. Fluctuations of these values are reported in Tables S1-S3 of the Supplementary Information.

Individual level resolution

The results presented so far are broken down into categories of participants and provide an aggregated quantitative estimation of the interaction behavior inter- and intra-classes. Additional information can however be gathered at the individual level, given the high resolution of the infrastructure used for the data collection. A reconstruction of the network of interactions among single individuals can be achieved to inspect in a deeper fashion the aggregated features reported above. Figure 6 shows a set of interaction networks for each pair of classes and within each class, corresponding to the entire monitoring period. In these networks, a node represents an individual, and an edge is drawn between two individuals whenever a face-to-face proximity event involving them was recorded. The networks restricted to physicians, nurses or ward assistants are rather dense, indicating a large diversity of contacts: within a given role, each health care worker interacts with many others. The picture is completely different for patients and caregivers: not only are there very few contacts between caregivers or between patients, but, as expected, the patient-caregiver contacts are very specific. Each patient has contacts with essentially one caregiver, and vice-versa, which corresponds to the fact that each patient was accompanied by one caregiver, confirming at the individual level the results previously observed at the class level. Contacts among caregivers and among patients are barely observed.

**Discussion**

Mathematical and computational models play an increasingly important role in the assessment and control of epidemics. In addition to studies focusing on the general population, models are also used to study the spread of infectious diseases in specific settings, such as, e.g., hospitals. Nosocomial infections represent a major public health issue with high morbidity and mortality and high costs associated with prolonged treatments, which deserve urgent and efficacious prevention strategies. The sources and transmission paths of nosocomial infections are indeed often unknown and unrecognized. The application of epidemic models to hospitals is therefore crucial to provide valuable insights on the routes of infection propagation and to identify tailored measures for prevention and control of hospital acquired infections. Models need to be informed with the pattern of interactions among individuals along which the transmission of infection can occur, however only simple homogeneous assumptions have been considered so far. Our study provides for the first time direct measures of the number and duration of close contacts by different role and at the individual level in a hospital setting. By taking advantage of the RFID technology, our results account for important heterogeneities in the hospital population and in the interactions among patients, health care workers, and visitors, that enable an accurate parameterization of models for infectious disease spread on the close-contact route.

Our main finding is the very limited interaction that we observed between pairs of patients or between pairs of caregivers, and between health care workers and caregivers. This is empirically found both in the number of contact events, taking into account both distinct and repeated events, and in the duration of such contacts. This result has immediate practical implications for the development of prevention measures for respiratory infections within the hospital, which represent the most frequent nosocomial infections (27,28). Current guidelines (18) identify the caregiver class as the priority group that control strategies should target (29), given that they may carry asymptomatic or mild community acquired respiratory illnesses and then spread the infection within the hospital to susceptible patients and staff. While caregivers may represent a source of the infection, our results show that their pattern of interaction is very stable and mostly spent in contact with the corresponding patient, thus not favoring the spread of a potential infection to a large number of individuals in the ward. This observation can inform models aimed at testing different prevention recommendations, exploring control resources focused mainly on the caregiver-patient interaction as opposed to resources focused on all possible interactions that a caregiver may have in the ward.

Another major insight concerns patients. Our results show that in addition to intense and continuative interactions with the caregivers, patients are contacted most frequently and with the longest duration by nurses, among all health care categories. Nurses were also found to have a pattern of frequent and long contacts among each other. Thus, in our setting, where a high proportion of patients had a diagnosis of respiratory infection, the intensive professional contact of nurses with patients and among themselves may result in a higher risk for airborne infections among nurses. These observations highlight the crucial importance of prioritizing nurses in local infection control interventions, and confirm the findings of a recent study where data on contacts were collected through a questionnaire (30). On the other hand, the pattern of contacts between physicians and patients showed a small number of contacts of short duration suggesting a less important role of this health care category in infection transmission. It is worth to mention that, although our study was conducted during the peak of influenza A/H1N1v activity in Italy, no case of influenza transmission was observed among the individuals included in the study. This observation is in line with a strict application of H1N1 containment guidelines in the hospital setting established by the Italian Ministry of Health (31).

We have studied the statistics associated with the collected data and we found a large degree of heterogeneity in the number of contacts and in the duration of contacts across classes. Large fluctuations up to 2 or 3 orders of magnitude are observed in the number of contacts established by an individual in a given class with any other individual, and similar results are observed in the duration of contacts. These fluctuations are however reduced when considering specific class-class interactions, highlighting the presence of well-defined interaction behaviors that are class-specific. Large variations are observed in the patient-caregiver interaction, providing additional empirical evidence for the importance of focusing control efforts on such interactions as they may lead to large variations in a potential outbreak.

Though observed in the hospital context, the large variability observed in the contact number and contact duration is consistent with empirical data collected in other settings. As an example, the same RFID technology was applied to investigate contact patterns in scientific conferences (23,25) showing also in this case a strong heterogeneity of the contact durations between individuals – most contacts were very short in this setting, but contacts of very different durations were observed, including very long ones. This similarity points to the presence of common statistical signatures in the way people interact, that go beyond the constraints and behavioral patterns imposed by the specific context.

Our study allows to define risks of transmission between classes of individuals by calculating matrices of contacts of class-class type along the three quantities describing a contact event that we considered – the number of contacts, the number of distinct contacts, the overall duration of contacts. These contact matrices represent the input ingredients for the parameterization of mathematical and computational models of nosocomial infections, going beyond simple homogeneous assumptions and simple structuring of the population into two classes – patients and health care workers – as in previous studies (20,32). This work complements similar efforts that focus on the community level (9,10), though our data collection method ensures a higher objectivity of the measure of contact events, and a higher resolution both in time and in space. By taking full advantage of this resolution, RFID technology can be used to inform increasingly complex models that require a finer classification of contacts by personal characteristics (33,34). Differently from survey methodologies, our method affords data collection and behavior characterization at the individual level, and thus may be used to inform agent-based modeling approaches. This could provide additional insights and uncover unexpected behaviors induced by the fluctuations observed even within the class-class structuring of the population.

However, in order to obtain reliable and statistically significant results from numerical simulations, longer and more extensive deployments should be considered to better characterize expected behaviors and the associated fluctuations. Although this study was carried out over a period of one week, it was conducted in a period when strict rules for infection control were applied because of the threat of cross infection with the 2009 A/H1N1 pandemic influenza virus. We do not believe that the conduction of the study itself has changed the behavior of health care workers, but longer deployments would in this respect be needed to assess the stability of the results over time.

A strength of the present study is the high participation rate of individuals in our experiment. We did not encounter major opposition in the acceptance of the experimental procedure by health care workers, patients, or caregivers, and only a very limited fraction of them (3.6%) declined participation. Once enrolled in the study, most participants regularly wore the RFID devices according to the instructions, thus contributing to the collection of high quality data. These observations suggest that if appropriate protocols are provided, and privacy protection is ensured, measures of contacts through RFID devices are easily replicable, and it is conceivable to aim at fully covering entire wards or hospitals. In this respect, such an approach would constitute a major improvement in the collection of high quality data, if compared to available studies based on interviews where the participation rate was much lower and where logistics and resources limited the length and coverage of the survey (9,28).

Our method presents limitations as well. First of all, given that the RFID tags exchange radio-frequency signals, the collected data can only provide information on the proximity of two badges (and therefore of the persons carrying the badges), but no information on the possible occurrence of a physical contact between the two persons is available. Our measures can thus be used for properly estimating the transmission parameters of respiratory infections but they are less informative for infections transmitted by direct contact. Note however that physical contact can only occur between persons who are already in spatial proximity. Therefore, it would be very interesting to study the fraction of close encounters that result in a physical contact, as this may help to identify useful parameters for modeling infections transmitted through physical contact. In the future, the use of sensors that can directly resolve physical contact may be explored. Another possible limitation is the fact that settings with no wireless connections involve additional complexity in the implementation of the measurements and in the definition of procedures, as limited by the available communication infrastructure. However, upcoming new technology that will allow operating the RFID sensing layer in a fully distributed fashion with on-board storage on the devices, will have minimal requirement on the host infrastructure and is expected to provide increasingly larger opportunities for deployment of wearable sensing systems for the measurement of contacts patterns. This new technology would be also very important to run similar data collection campaigns that cover full hospitals, to assess whether logistic and behavioral characteristics and procedures that are ward-specific are reflected in the observed contact patterns. Finally, a comparative study between different hospitals may provide valuable insights into how structural organization and procedural management may impact the contact patterns among individuals, and therefore the potential epidemic spreading within the hospital. This would also allow us to assess the specificity of the results obtained in the present study. On the other hand, human behavior has been shown in many studies to exhibit important regularities, and a certain number of characteristics of our results can be expected to hold across different wards or hospitals: for example the broad distributions of Fig. 3, or the overall structure of the interaction networks, with the strong specificity of patient-caregiver contacts contrasted to the HCW interactions.

Our study represents, to our knowledge, the first example of unsupervised data collection of face-to-face contacts in a hospital setting by means of wearable radio frequency devices. The obtained results provide significant advances in our knowledge of the mixing patterns taking place in a hospital ward, and allow for a fine structuring of the population into classes of individuals based on their role, along with the evaluation of the corresponding contact matrices. The resulting analysis may help to identify specific interactions at increased risk of transmission, and to explore a variety of possible interventions by means of numerical simulations obtained with modeling approaches informed by measured contact matrices. Outbreak investigations conducted concurrently with proximity sensing by wearable devices may further augment the knowledge we have on the routes of transmission and thus help in reducing the burden of nosocomial infection.

**Acknowledgments**

We warmly thank Bitmanufaktur and the OpenBeacon project. We also thank the health care workers who participated in the study and helped at the Bambino Gesù Hospital, the Head of the Department of Pediatrics, Prof. A.G. Ugazio, the personnel from the Information System Department, their Head, Eng. Giulio Siccardi, and the families of participating patients. This work has been partially supported by the Italian Ministry of Health (Grant n. 1M22) to C.R., A.T and C.C.; by the EC FET-Open project DYNANETS (contract number 233847) to L.I. and V.C.; and by the ERC Ideas grant EpiFor (contract no. ERC-2007-Stg204863) to V.C. The funders had no role in study design, data collection and analysis, decision to publish, or preparation of the manuscript.

**Author Contributions**

A.E. Tozzi, C. Cattuto, A. Barrat, C. Rizzo, and V. Colizza designed the study; data collection, on-site quality monitoring and logistics were managed by M. Romano, F. Gesualdo, E. Pandolfi, L. Ravà, C. Cattuto, W. Van de Broeck; L. Isella, C. Cattuto, A. Barrat, V. Colizza, and A.E. Tozzi performed the data analysis; A.E. Tozzi, C. Cattuto, C. Rizzo, L. Isella, V. Colizza, and A. Barrat wrote the manuscript; all the authors participated in the discussion on data findings.

**Competing Interests**

The authors declare no competing interest.

**Figures**

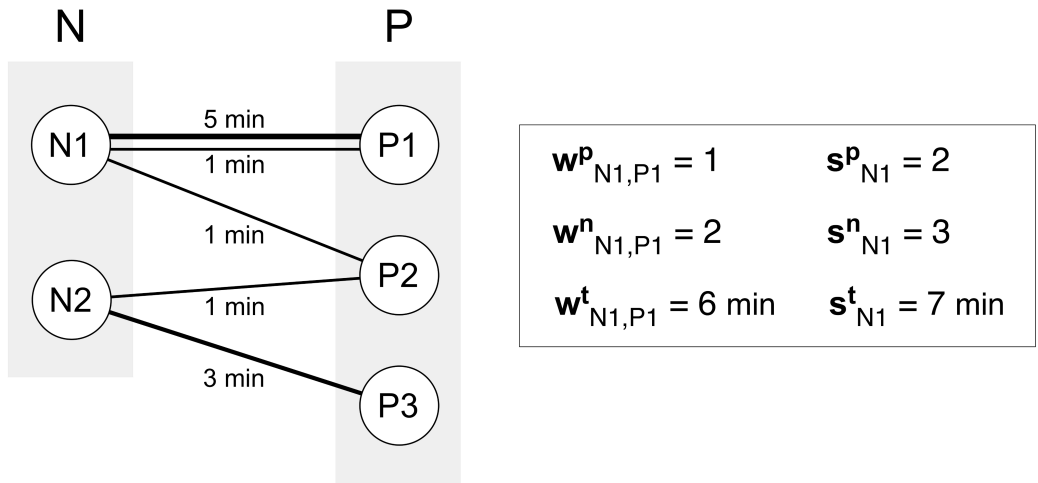

**Figure 1.** Schematic representation of detected contacts among 2 nurses (N1, N2) and 3 patients (P1, P2, P3) and corresponding measured quantities. Each individual is represented by a node and a link corresponds to a contact established between two individuals. The width of the link is a measure of the duration of the contact, also indicated explicitly in terms of minutes. Multiple links can occur between two individuals, as highlighted in the pair N1-P1, indicating a contact of frequency larger than 1. The quantities introduced in the Materials and Methods section are calculated for the pair of individuals N1 and P1. The pair established one contact ($w_{ij}^p = 1$) with frequency equal to two ($w_{ij}^n = 2$) for a total duration of six minutes ($w_{ij}^t = 6 \min$). By taking into account all interactions, individual N1 has established three contacts ($s_i^n = 3$), two of which were distinct contacts ($s_i^p = 2$), for a total duration of contacts equal to seven minutes ($s_i^t = 7 \min$).

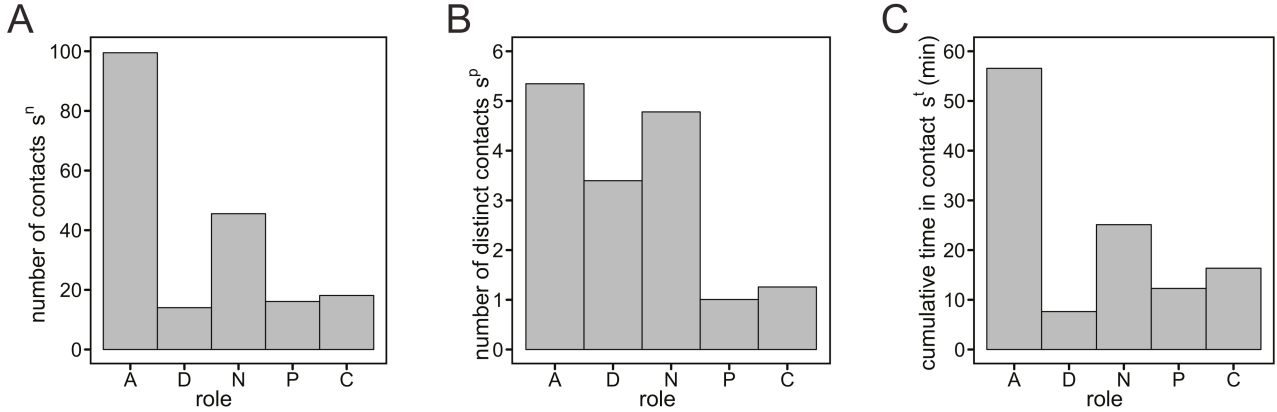

**Figure 2.** Contact number and duration per role (A: ward assistants; D: Doctor; N: Nurse; P: Patients; C: Caregiver). The plots show the median values per participant in each class of: the number of contacts $s^n$ (panel A), the number of distinct contacts $s^p$ (panel B), the cumulative time in contact $s^t$ (panel C). All quantities for a given class are computed on the contacts established by participants in that class with any other participant. Data are normalized to a 24-hour interval.

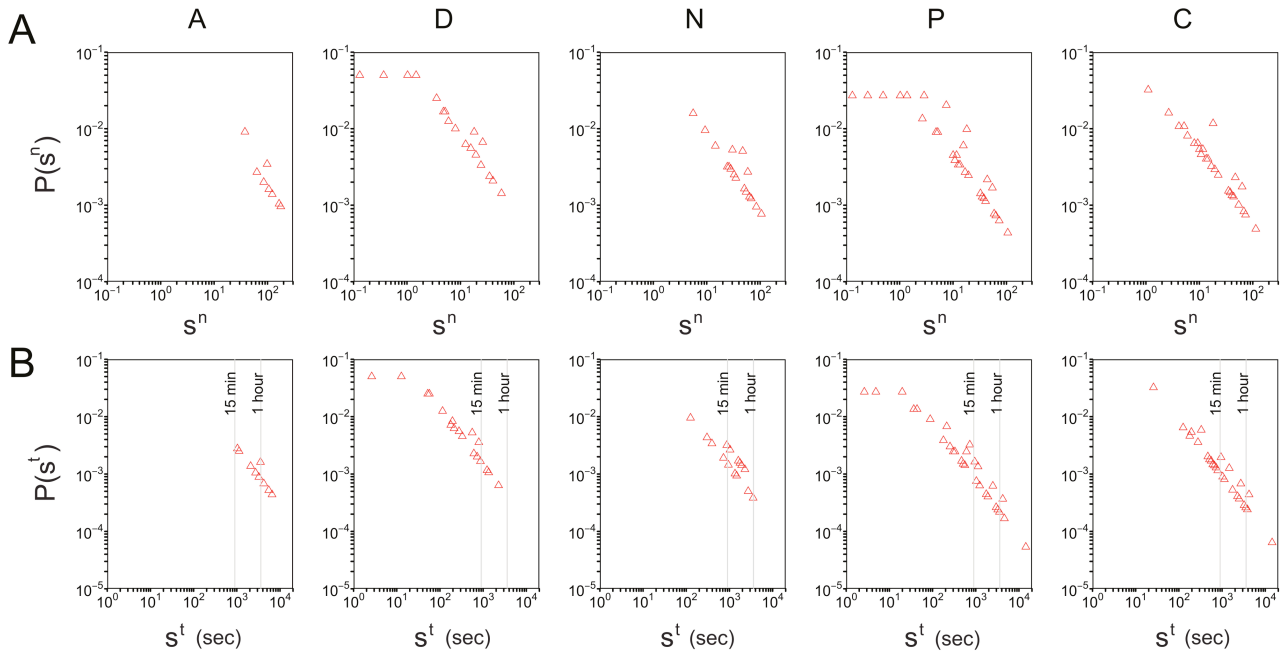

**Figure 3.** Probability density functions of the number of contacts per individual, $s^n$ (panel A), and of the cumulative time in contact $s^t$ (panel B). Each plot corresponds to a given class and considers the contacts that an individual in that class established with any other individual. Contact duration is expressed in seconds and is normalized to a 24-hour interval.

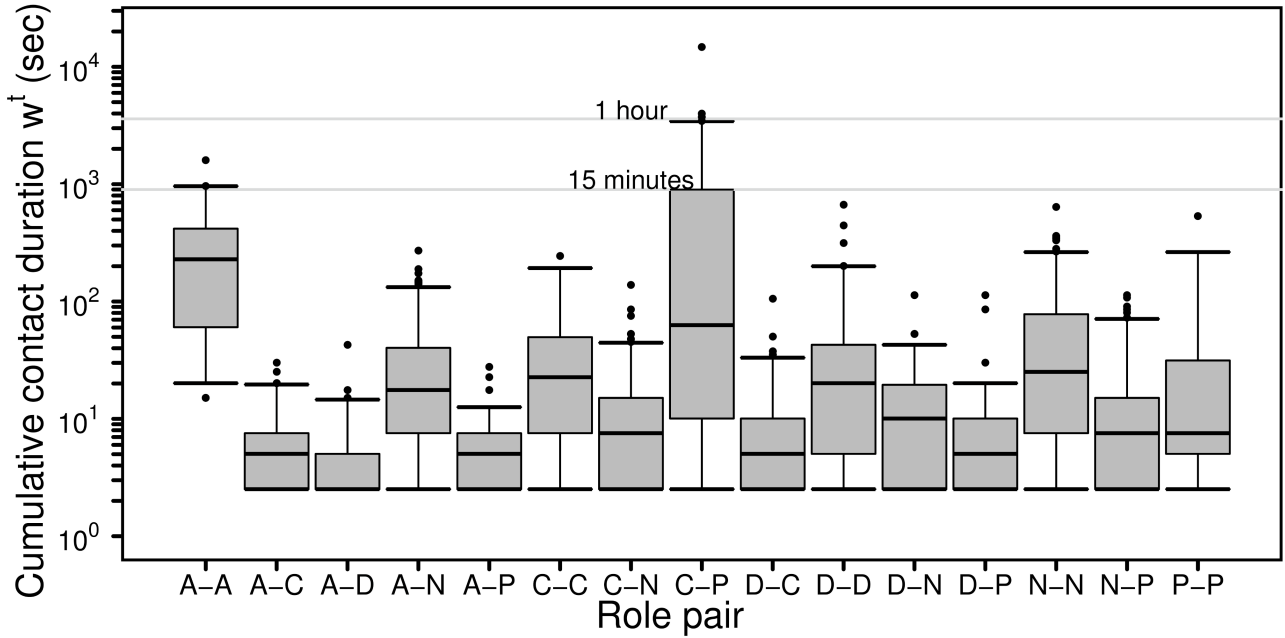

**Figure 4.** Boxplots for the distributions of cumulative contact durations $w^t$ between individuals belonging to given role pairs (horizontal axis), given the occurrence of a contact. Here we only consider non-zero values of $w^t$, and contact durations are expressed in seconds and are normalized to a 24-hour interval. On normalizing, the experimental resolution of 20 seconds yields the lowest value of 2.5 seconds visible in the figure. As usual, the bottom and top of the boxes correspond to the 25$^{th}$ and 75$^{th}$ percentiles, and the horizontal segment indicates the median. The ends of the whiskers correspond to the 5$^{th}$ and 95$^{th}$ percentiles. The dots are outliers located outside the 90% confidence interval, i.e., events falling below the 5$^{th}$ percentile or above the 95$^{th}$ percentile.

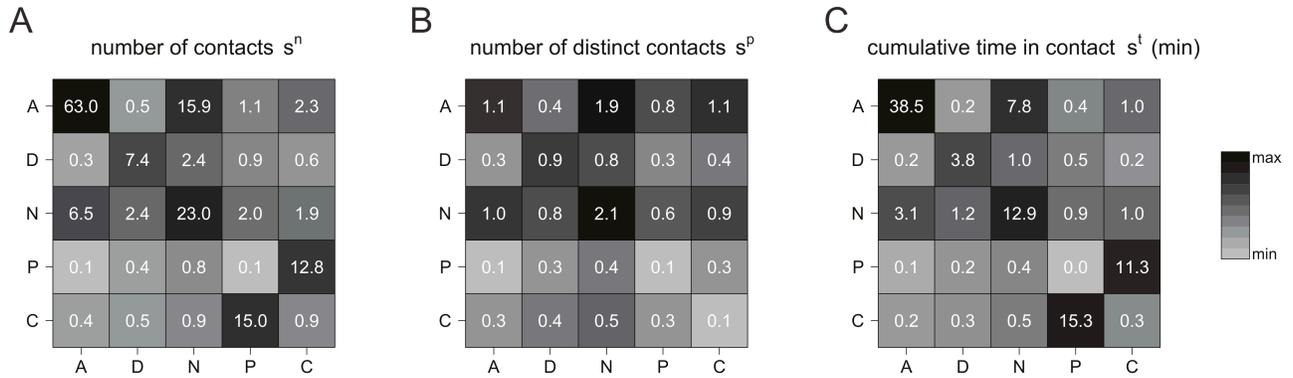

**Figure 5.** Contact matrices defined on the classes of individuals. Matrices are displayed for the number of contacts $s^n$ (panel A), the number of distinct contacts $s^p$ (panel B), and the cumulative time in contact $s^t$ (panel C). The matrix entry for classes X (row) and Y (column) is the median value of the node strengths for individuals of class X, computed on the contacts they had with individuals of class Y; the asymmetry of the matrices depends on the different numbers of individuals populating each class (3). Individuals of class X that did not have contacts with individuals of class Y count as nodes with zero strength, i.e., they affect the median value for the corresponding matrix entry. To increase the readability of the figure, matrix entries are grayscale-coded according to the median values, with the lightest and darkest shade of gray respectively corresponding to the minimum and maximum value for each matrix. Contact durations are expressed in minutes and normalized to a 24-hour interval.

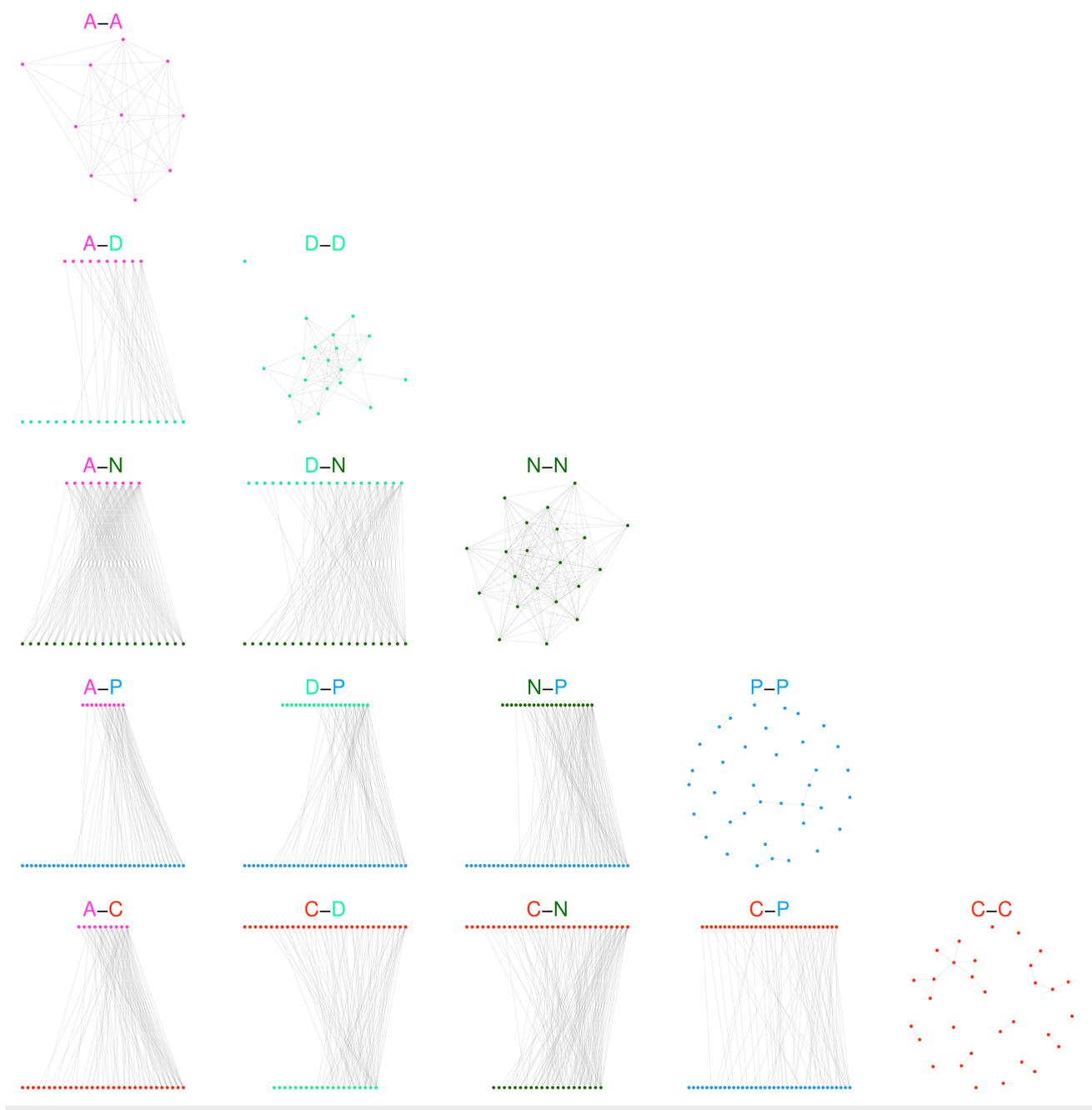

**Figure 6.** Cumulative contact networks of individuals, for all pairs of classes and within each class. Nodes represent unique individuals, and edges between nodes represent a cumulative face-to-face time over the whole monitoring period. In the off-diagonal layouts, nodes are positioned from left to right in increasing order of number of edges.

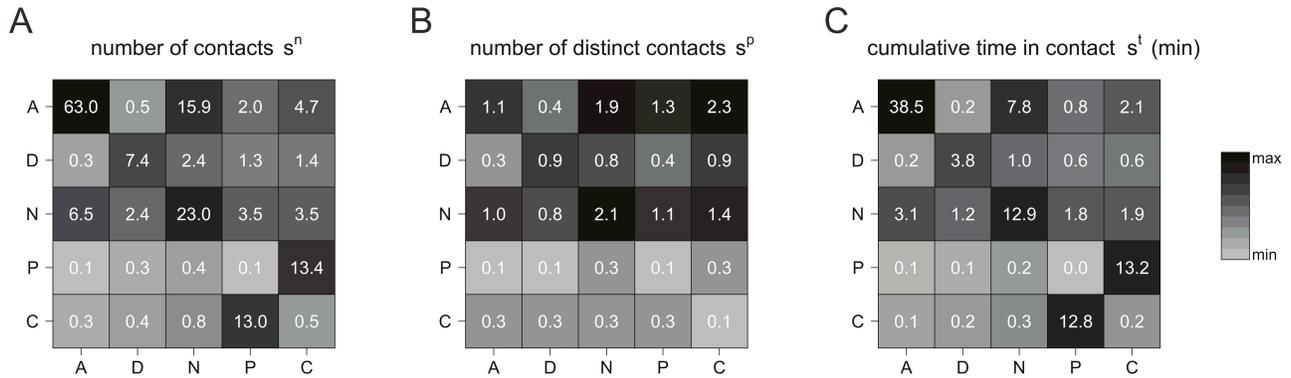

**Figure S1.** Contact matrices for classes of individuals, with no filtering of RFID badges. The matrices are computed in the same way as those of Figure 5, but no filtering procedure is applied and the data for all RFID badges are retained. Matrices are displayed for the number of contacts $s^n$ (panel A), the number of distinct contacts $s^p$ (panel B), and the cumulative time in contact $s^t$ (panel C). Matrix entries are grayscale-coded according to the median values, with the lightest and darkest shade of gray respectively corresponding to the minimum and maximum value for each matrix. Contact durations are expressed in minutes and normalized to a 24-hour interval. Comparison with Figure 5 shows the robustness of the data with respect to the filtering procedure.

**Tables**

**Table 1.** Characteristics of patients included and excluded from the analysis.

|  | Patients excluded from analysis due to poor signal quality | Patients included in the analysis | Total |
|---|---|---|---|
| Number | 39 | 37 | 76 |
| Age, years; median (range) | 3.04 (41 d – 17 y) | 3.54 (44 d – 17 y) | 3.38 (41 d – 17 y) |
| Length of stay, days; median (range) | 5 (1-36) | 7 (1-29) | 6 (1-36) |
| No. of patients with H1N1 infection (%) | 7 (17.9%) | 12 (32.4%) | 19 (25.0%) |
| No. of patients with acute respiratory infections other than H1N1 (%) | 18 (46.1%) | 14 (37.8%) | 32 (42.1%) |

**Table 2.** Characteristics of the study sample in terms of classes and number of contacts. The total number of daily contacts measured per class is the sum of all contacts $s^n$ established by individuals in that class. The number of daily contacts per participant is the median value of the number of contacts $s_i^n$, with *i* belonging to the given class.

| Role | No. participants (%) | Total no. daily contacts | No. daily contacts per participant [90% confidence interval] [90% CI] |
|---|---|---|---|
| (A) ward assistants | 10 (8.4 %) | 991.1 | 99.5 [38.3-172.8] |
| (N) Nurses | 21 (17.6 %) | 920.2 | 45.5 [9.2-83.1] |
| (C) Caregivers/ Accompanying persons/Visitors | 31 (26.1%) | 910.2 | 18.1 [3.4-69.4] |
| (P) Patients | 37 (31.1%) | 880.6 | 16.1 [0.5-63.6] |
| (D) Physicians | 20 (16.8%) | 325 | 14.0 [0.4-41.8] |